\DocumentMetadata{}

\documentclass[manuscript]{acmart}

\usepackage[utf8]{inputenc}
\usepackage{amsthm}
\usepackage{amsmath,tikz-cd}
\usepackage{natbib}
\usepackage{booktabs}
\usepackage{blindtext}
\usepackage{graphicx}
\usepackage{float}
\usepackage{hyperref}
\usepackage[toc]{appendix}
\usepackage[normalem]{ulem}
\usepackage{algorithm}
\usepackage{algpseudocode}
\usepackage[font=small,skip=0pt]{caption}
\usepackage[font=small,skip=0pt]{subcaption}
\usepackage{multirow}
\usepackage{array}
\setkeys{Gin}{width=\textwidth,keepaspectratio}

\newtheorem{cor}{Corollary}[subsection]

\newtheorem{defn}[cor]{Definition}

\theoremstyle{remark}

\begin{document}

\title{Empirical Power Analysis of a Statistical Test to Quantify Gerrymandering}

\author[R.A.Clark]{Ranthony A. Clark}
\email{ranthony.clark@duke.edu}
\affiliation{%
  \institution{Department of Mathematics, Duke University}
  \city{Durham}
  \state{NC}
  \country{USA}
}

\author[S. Glenn]{Susan Glenn}
\email{sglenn@wisc.edu}
\affiliation{%
  \institution{Los Alamos National Laboratory}
\city{Los Alamos}
  \state{NM}
  \country{USA}
}

\author[H. Lee]{Harlin Lee}
\email{harlin@unc.edu}
\affiliation{%
  \institution{School of Data Science and Society, University of North Carolina at Chapel Hill}
    \city{Chapel Hill}
  \state{NC}
  \country{USA}
}

\author[S. Villar]{Soledad Villar}
\email{soledad.villar@jhu.edu}
\affiliation{%
  \institution{Department of Applied Mathematics and Statistics, Johns Hopkins University}
    \city{Baltimore}
  \state{MD}
  \country{USA}
}

\newcommand{\change}[1]{\textcolor{blue}{#1}}

\begin{abstract}  
Gerrymandering is a pervasive problem within the US political system. In the past decade, methods based on Markov Chain Monte Carlo (MCMC) sampling and statistical outlier tests have been proposed to quantify gerrymandering and were used as evidence in several high-profile legal cases. We perform an empirical power analysis of one such hypothesis test from Chikina et al (2020). We generate a family of biased North Carolina congressional district maps using the 2012 and 2016 presidential elections and assess under which conditions the outlier test fails to flag them at the specified Type I error level. The power of the outlier test is found to be relatively stable across political parties, election years, lengths of the MCMC chain and effect sizes. The main effect on the power of the test is shown to be the choice of the bias metric. This is the first work that computationally verifies the power of statistical tests used in gerrymandering cases. 
\end{abstract}

\maketitle

\tableofcontents

\section{Introduction}

\begin{figure}[htbp]
    \centering
    \begin{minipage}[b]{0.45\textwidth}
        \centering
        \includegraphics[width=\textwidth]{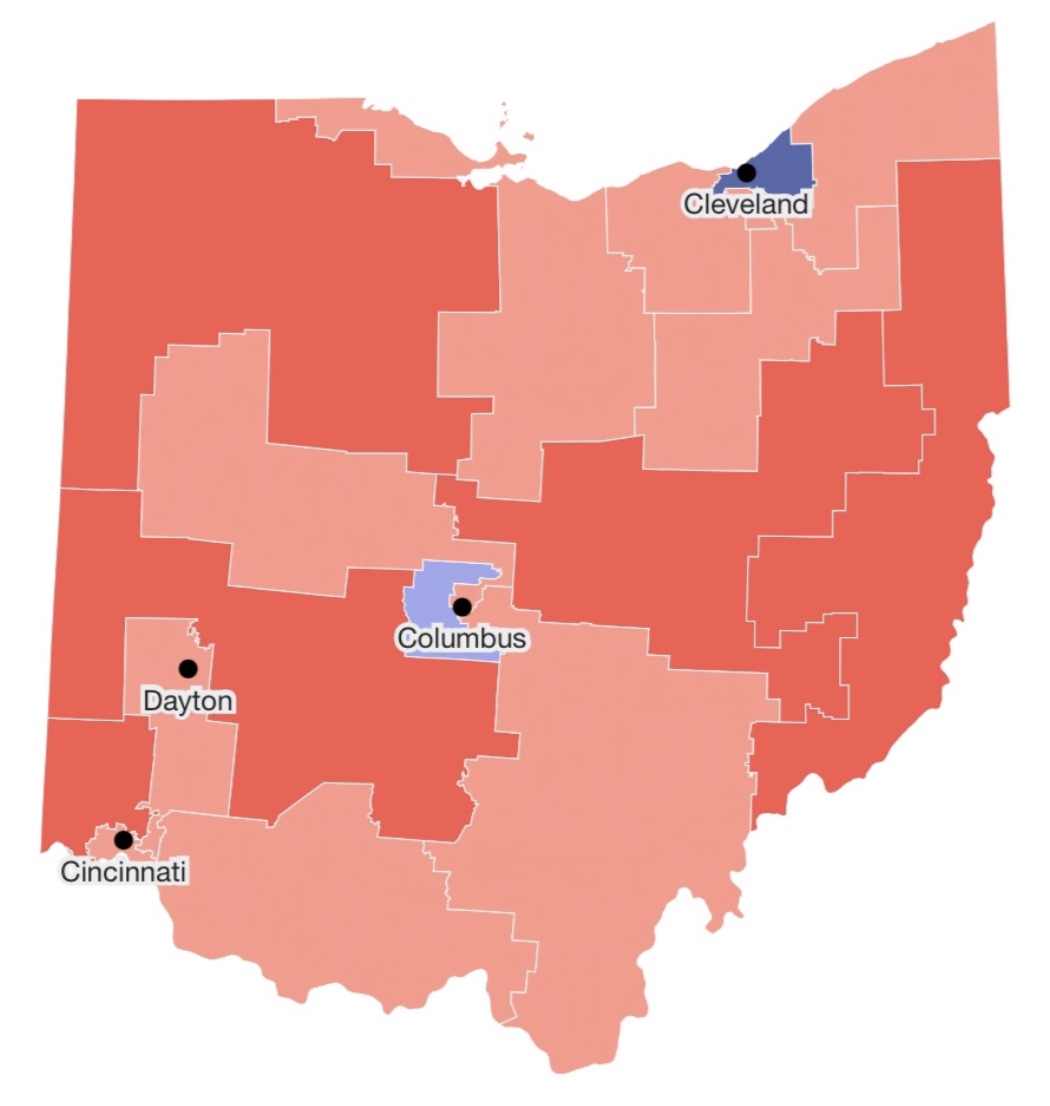} 
        \caption*{(a) 2021 Ohio Congressional Districting Plan (Ohio Redistricting Commission).}
        \label{fig:ohio-cd}
    \end{minipage}
    \hfill
    \begin{minipage}[b]{0.45\textwidth}
        \centering
        \includegraphics[width=\textwidth]{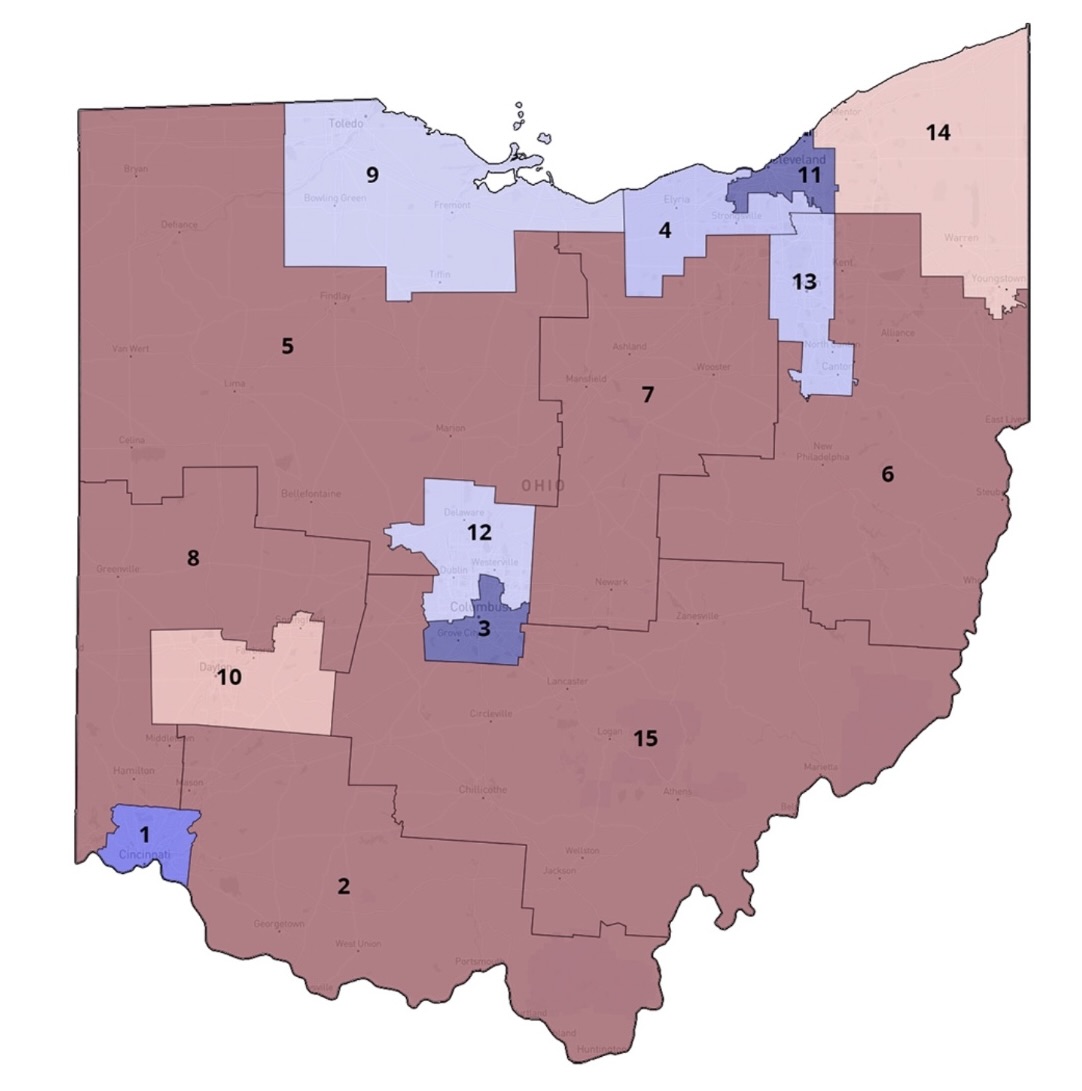} 
        \caption*{(b) 2021 Ohio Congressional Districting Plan (Ohio Citizens' Redistricting Commission).}
        \label{fig:ohio-cccd}
    \end{minipage}
    \vspace{1em} 
    \caption{This figures shows two possible redistricting plans for the state of Ohio in 2021. The plan (left) by the Ohio Redistricting Commission features thirteen Republican leaning districts and two Democrat leaning districts, while the plan (right) proposed by the Ohio Citizens' Redistricting Commission features eight Republican leaning districts and seven Democrat leaning districts. Ohio has 15 congressional districts and a statewide vote share of $\sim$54\% Republican and $\sim$46\% Democrat. Given the same distribution of votes, each map produces very different outcomes in representation. }
    \Description[2021 Ohio Congressional Districting Plans]{2021 Ohio Congressional Districting Plans}
    \label{fig:ohio-map-matters}
\end{figure}

Redistricting is the process by which every ten years, a state is partitioned into zones, called districts, in which elections are held to select representatives. The number of districts in each state is determined by an apportionment formula related to a state's increasing or decreasing population according to the decennial US Census. For example, the state of North Carolina has gained an additional congressional district after each of the most recent redistricting cycles, starting with 12 congressional districts in 2000, increasing to 13 in 2010, and 14 in 2020. A great deal of attention is paid to how these districts are drawn, as the same distribution of votes can lead to dramatically different outcomes in representation depending on the choice of map (See Figure \ref{fig:ohio-map-matters}). The practice of manipulating the lines for redistricting in such a way that overtly advantages a particular party or class of people is called \emph{gerrymandering}. Given the stakes involved in ensuring fair representation, while much research has been done by political scientists on redistricting, there has been significant interest over the past two decades by mathematicians and other computational scientists to quantify gerrymandering--that is to utilize mathematical, statistical, and computational tools to study notions of fairness broadly construed in electoral redistricting. 

In computational redistricting, the process of dividing a state into districts is frequently viewed as a graph partitioning problem, where a partition of a graph into $d$ subgraphs subject to certain constraints models the partitioning of a state into $d$ districts. While it is tempting to consider a discrete approach where all possible $d$-partitions of a state are enumerated and then one optimizes towards a map that is `best' with respect to a given state's policies and priorities, the number of possible redistricting plans for a given state exceeds the number of atoms in the universe. To mitigate this phenomenon, practitioners utilize quantitative methodologies based on Markov Chain Monte Carlo (MCMC) to sample from the large universe of possible redistricting plans \cite{fifield2020automated,CANNON202475,cannon2023voting,herschlag2020quantifying,deford2019recombination}. A dominant methodology in assessing the fairness of a particular plan is referred to as \emph{ensemble analysis}, where MCMC sampling is used to generate a large collection, or ensemble, of maps satisfying certain non-partisan constraints for a valid redistricting plan. This ensemble can then serve as a probative baseline of comparison for a given map that may be subject to partisan bias.

One drawback of this methodology is that in the application of political redistricting, as opposed to theoretical work in MCMC, it is rarely possible (or even desirable) to compute the mixing time of a chain. Instead, researchers in this domain focus on convergence heuristics and sensitivity analysis to support replicability and confidence in their findings \cite{deford2019recombination}. Recent work by DeFord, Moon and Solomon devised and implemented Gerry Chain: an open source software that produces more compact maps and empirically seem to mix faster than prior ones \cite{gerrychain}. However, it is still unclear how to assess that the Markov Chain has mixed, and that the sampling is actually being performed from the stationary distribution. In \emph{``Assessing significance in a markov chain without mixing''} by Chikina et al \cite{chikina2017assessing} they develop a local statistical test that provides a bound for the probability that a given map was chosen from the uniform distribution without the assumption of mixing  \cite{deford2019recombination} by providing a p-value. This gives a rigorous way to detect whether a given map has been `carefully crafted' in its partisan bias. 

In practice, this work was used to provide expert witness testimony in the 2018 case \emph{League of Women Voters of Pennsylvania v. Commonwealth of Pennsylvania}, which ruled that the 2011 Pennsylvania congressional districting plan was an unconstitutional gerrymander \cite{League_of_Women_Voters_v_Commonwealth}. While \emph{Rucho v. Common Cause} ruled that partisan gerrymandering was not federally judiciable, there remain many ongoing cases at the state level related to partisan gerrymandering. Given the impact of this test in legislation, it is of interest to understand its strength and limitations. Is it possible to `game' the test so it does not detect maps with a certain type of partisan bias as gerrymanders? 

\subsection{Statistical Tests for Gerrymandering}\label{sec:StatisticalTest}

The goal of this current project is to empirically assess the power of statistical outlier tests with a view towards the domain application of gerrymandering in computational redistricting. Our focus is on the outlier test (Theorem \ref{thm:cfptest} below) introduced in \cite{chikina2017assessing} and later improved upon in  \cite{chikina2020separating} (Theorem \ref{thm:eatest} below). 

Let $\mathcal{M}$ be a reversible Markov chain on a state space $\Sigma$ along with an associated label function $\omega: \Sigma \rightarrow \mathbb{R}$. If $\sigma_0$ is some given state, the $\epsilon$-outlier test gives a measure of how unusual $\sigma_0$ is compared to other states chosen randomly from the stationary distribution $\pi$ of $\Sigma$. In redistricting we take $\Sigma$ to be the space of all valid congressional redistricting plans. By valid we mean maps that satisfy federal and state criteria for population balance, contiguity, and compactness, where the latter two are related to the connectedness and the regularity of the shape of the districts respectively. The label function $\omega$ is a metric that ascribes to a given plan some measure of its bias, which we will generally refer to as partisan metrics. For an initial state, or redistricting plan, we can compare its label to those in a trajectory coming from the intial plan via some Markov chain algorithm like those described in Section ~\ref{subsec:alg}.

If we consider a trajectory $\sigma_0, \sigma_1, \hdots, \sigma_k$ and the associated values of the label function \\ $\omega(\sigma_0), \omega(\sigma_1),\hdots, \omega(\sigma_k)$ for some fixed $k$ steps in the chain, then $\sigma_0$ is an \emph{$\epsilon$-outlier} with respect to the states $\sigma_i$ with $i \in \{0, \hdots k\}$ if there are at most $\epsilon(k+1)$ indices $i$ such that $\omega(\sigma_i) \leq \omega(\sigma_0)$. This provides a quantitative framework to indicate that a redistricting plan that is an $\epsilon$-outlier with some probability compared to other local plans in the state space of a Markov chain is a partisan gerrymander. The following hypothesis test by Chikini, Frieze, and Pegden \cite{chikina2017assessing} says that $\omega(\sigma_0)$ is an $\epsilon$-outlier at significance $\sqrt{2\epsilon}$.

\begin{theorem}\label{thm:cfptest}[CFP test \cite{chikina2017assessing}] Let $\mathcal{M} = X_0, X_1, \hdots$ be a reversible Markov chain with a stationary distribution $\pi$, and suppose the states of $\mathcal{M}$ have real valued labels. If $X_0 \sim \pi$, then for any fixed $k$, the probability that the label of $X_0$ is an $\epsilon$-outlier from among the list of labels observed in the trajectory $X_0, X_1, X_2, \hdots, X_k$ is at most $\sqrt{2\epsilon}$.
\end{theorem}

Observe that in Theorem \ref{thm:cfptest}, both $\sigma_0$ as a random state in the chain $\mathcal{M}$ and the random trajectory from $\sigma_0$ are considered in the computation of the p-value. In subsequent work, Chikina, Frieze, Mattingly, and Pegden introduced another powerful test which improved interpretation of the test in Theorem \ref{thm:cfptest} by separating ``the measure of statistical significance from the question of the magnitude of the effect" \cite{chikina2020separating}. A key distinction between Theorem \ref{thm:cfptest} and Theorem \ref{thm:eatest} below is that for the former, the null hypothesis is that $\sigma_0 \sim \pi$ where as for the latter the null hypothesis is that $\sigma_0$ is not an $(\epsilon,\alpha)$-outlier. We now define an $(\epsilon,\alpha)$-outlier.  

\begin{defn}[($\epsilon,\alpha$)-outlier \cite{chikina2020separating}] \label{def:ea} 
Let $p_{0,\epsilon}^k(\sigma_0)$ be the probability that $\omega(\sigma_0)$ is among the smallest $\epsilon$ fraction of $\omega(X_0),\omega(X_1), \hdots, \omega(X_k)$ where $\sigma_0=X_0, X_1, \hdots, X_k$ is a trajectory. Then with respect to $k$, the state $\sigma_0$ is an \emph{$(\epsilon,\alpha)$-outlier} in $\mathcal{M}$ if, among all states in $\mathcal{M}$, $p_{0,\epsilon}^k(\sigma_0)$ is in the largest $\alpha$ fraction of the values of $p_{0,\epsilon}^k(\sigma_0)$ over all states $\sigma \in \mathcal{M}$, weighted according to the stationary distribution $\pi$. 
\end{defn}

We will focus on empirically evaluating the accuracy of the following test in Section ~\ref{sec:EmpiricalResults}.  

\begin{theorem}\label{thm:eatest}[CFMP Test \cite{chikina2020separating}]
Consider $m$ independent trajectories
\begin{align*}
\mathcal{T}^1 & = (X_0^1, X_1^1, \hdots, X_k^1),\\
& \vdots  \\
\mathcal{T}^m & = (X_0^m, X_1^m, \hdots, X_k^m)
\end{align*}
of length $k$ in the reversible Markov chain $\mathcal{M}$ from a common starting point $X_0^1= \cdots = X_0^m = \sigma_0$. Define the random variable $\rho$ to be the number of trajectories $\mathcal{T}^i$ on which $\sigma_0$ is an $\epsilon$-outlier. If $\sigma_0$ is not an $(\epsilon,\alpha)$-outlier, then 

\begin{align}\label{eq:NullDistribution}
\text{Pr}\left(\rho \geq m\sqrt{\frac{2\epsilon}{\alpha}}+r\right) \leq e^{-\min\left(\frac{r^2\sqrt{\alpha/2\epsilon}}{3m}, \frac{r}{3}\right)}.
\end{align}

\end{theorem}

Note that there are four parameters which will directly effect the power of the hypothesis test: $k, m, \alpha, \varepsilon$. As the number of trajectories ($m$) or the number of steps in each trajectory ($k$) increase, the test becomes more powerful. The effect size ($\varepsilon$) is the parameter which determines how extreme the outlier should be for the null hypothesis to be rejected. Therefore, a smaller $\varepsilon$ would mean the likelihood of rejecting the null is smaller, and thus the test is less powerful. Similarly, the significance level $\alpha$ is the Type I error rate for the test, where larger $\alpha$ values lead to higher power. The tail probability of the null distribution is upper-bounded by the expression on the right hand side of Equation~\ref{eq:NullDistribution} which means that in practice, the Type I error will never exceed $\alpha$. For the test to be non-trivial, $\varepsilon \geq 1/k $ and $2 \varepsilon < \alpha$.

\subsection{Contributions}

The goal of this paper is to investigate the local outlier test (Theorem ~\ref{thm:eatest}) in its ability to detect plans with extreme partisan bias. We provide a novel contribution as the first work that computationally verifies the power of statistical tests used in gerrymandering cases. This analysis and our pipeline to generate plans to probe this test are the primary contributions of this work: \\

\noindent\textbf{Highlighted Contributions}
\begin{enumerate}
    \item Empirical methodology for generating biased maps with respect to different partisan metrics. 
    \item Empirical power analysis of statistical outlier tests for gerrymandering.\\
\end{enumerate}

With regards to (1), note that we provide replicable algorithms utilizing open source software to generate maps that are intentionally ‘gerrymandered’ with respect to certain metrics, providing collections of maps that can be used as baselines for comparison for suspected biased plans. In addition, generating plans that are extreme with respect to certain partisan aims is of general interest in computational redistricting. \cite{cannon2023voting} recently introduced short bursts as a way to maximize the number of majority minority districts in a given state, and it is an open question how this method of generating ensembles behaves with respect to a variety of metrics, and also how
the choice of metric to be optimized interacts with the burst length. This has been considered by once to our knowledge in \cite{ratliff2024don} and referenced in \cite{Veomett2024}, but this is the first application with the general aim of `generating gerrymanders.' 

Our algorithms successfully produce biased ensembles with respect to select partisan metrics. In many cases the hill climbing method and short bursts methods produce similar levels of bias over the entire run of the chain, but short bursts often achieved those extreme scores in fewer steps. Lastly, given the
numerous metrics that exists to detect partisan gerrymandering, our work highlights
important trade offs and interactions of some of these standard metrics with one another (see Figure ~\ref{fig:Interactionplots}), contributing to literature that has investigating the interplay of these metrics \cite{deford2021implementing, bernstein2022measuring, katz2020theoretical, Veomett2024, ratliff2024don}.

For (2), we determine that the power of the outlier test is found to be relatively stable across political parties, election years, lengths of the MCMC chain and effect sizes. The main effect on the power of the test is shown to be the choice of the bias metric.

\subsection{Overview of Paper}

In Section ~\ref{sec:redistricting} we provide an overview of the relevant redistricting background for this work. There Section ~\ref{subsec:alg} describes the algorithms Flip Node and Recombination used for ensemble analysis in computational redistricting while Section ~\ref{subsec:partisan} provides an overview of several metrics that can be utilized to assess how partisan a particular redistricting plan is. 

In Section 3 we review our metholodgy for generating biased ensembles of plans in North Carolina for the 2012 and 2016 presidential elections and discuss the performance of the hill climbing method versus the short burst method. Finally, in Section 4 we do the empirical power analysis for the statistical test in Theorem ~\ref{thm:eatest} and conclude with a brief discussion of that analysis in Section 5.

\section{Redistricting Background}\label{sec:redistricting}

\subsection{Redistricting as a Graph Partition Problem}

We are concerned with generating and analyzing large collections of valid redistricting plans for a given state. Given a graph $G=(V,E)$ of the geography of a state (where the nodes in $V$ correspond to units such as precincts, census blocks, etc. across the state), a valid redistricting is a $d$-partition $V_1 \sqcup \hdots \sqcup V_d$ of $G$ into subgraphs subject to certain non partisan constraints. In this context, we operationalize population balance, contiguity, and compactness. For population balance we require that the sum of the population of the nodes in each subgraph are within 2\% of the ideal district population, which is given by the total population across the state divided by the number of districts. Below we provide general formulation of bounded restraints for a given district where $p:V\to \mathbb{R}$ is a population function on each node and $p(V_i)$ and $p(G)$ is the sum of this function over all vertices in a district and over the graph respectively,
\[
(1-\epsilon) \cdot \frac{p(G)}{d} \leq p(V_i) \leq (1+\epsilon) \cdot \frac{p(G)}{d}.
\]
For contiguity, we require that each subgraph $V_i$ is connected. Lastly, for compactness, we utilize what are called cut edges, where an edge is cut if its endpoints belong to different districts. If we let $f:V \to \mathbb{R}$ be such that $f(v)=i$ if $v \in V_i$, then we denote the set of cut edges, denoted $C$, of $G$ by 
\[
C = \{(v_1,v_2) \in E \mid f(v_1) \neq f(v_2)\}.
\]

The cardinality of $C$ will be smaller for plans that are more compact.

\subsection{Algorithms for Generating Districting Plans}\label{subsec:alg} 

In order to generate a new graph partition, i.e. redistricting plan, from an existing one, we can take a simple walk on $G$ via some MCMC algorithm. Here we focus on two called Flip node and Recombination. 

\subsubsection{Flip node}

A natural way to generate states in a Markov chain when each state is a redistricting plan modeled as a graph partition, is to flip nodes between bordering districts in such a way the preserves the contiguity of the plan. In the Flip node algorithm, a single node is chosen whose endpoint is among the edges in $C$. In order to account for the variability in the degrees of such nodes within the graph, \cite{chikina2017assessing} sampled uniformly from the set of partitions that only differ by the position of a single node on the boundary of a district, that is the set of pairs $(v,V_i)$ with the node $v \in C$ so that there exists a $e=(v,w)\in C$ with $f(w)=V_i$ \cite{deford2019recombination}. This formulation of the Flip node algorithm creates a reversible Markov chain.

\subsubsection{Recombination}

One widely utilized MCMC algorithm in redistricting is called Recombination (Recom), inspired by a similar natural recombination in biology. It works by starting with a given partition, randomly selecting two adjacent districts $V_i$ and $V_j$, merging them together, and drawing a minimal spanning tree on the nodes comprising $V_i \cup V_j$. By a \emph{spanning tree} we mean a subgraph with no cycles containing all $n$ nodes of the two merged districts with $n-1$ edges. This spanning tree can be cut to form two new districts $V'_i$ and $V'_j$. A proposal step then determines if these two new districts are accepted based on whether they satisfy constraints related to compactness and population balance. If accepted then the next state of the Markov chain is the new redistricting plan $V_1 \sqcup \hdots \sqcup V'_i \sqcup V'_j \hdots \sqcup V_d$, and if rejected, a new spanning tree is drawn and the process repeats. 

\begin{figure}[htp]
\begin{subfigure}{\linewidth}
\centering
    \includegraphics[width=0.75\linewidth]{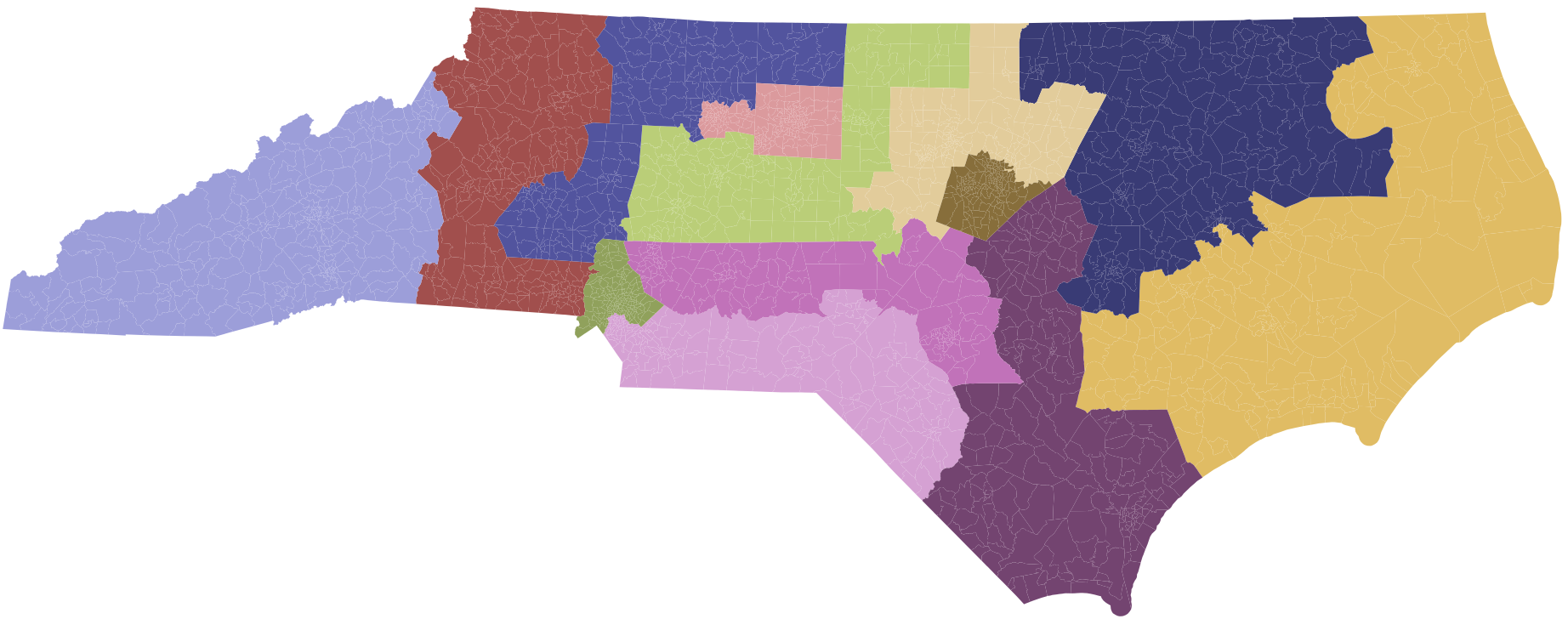}
    \caption{2020 NC congressional districting map}\label{fig:original_map}
\end{subfigure}
\begin{subfigure}{\linewidth}
\centering
    \includegraphics[width=0.75\linewidth]{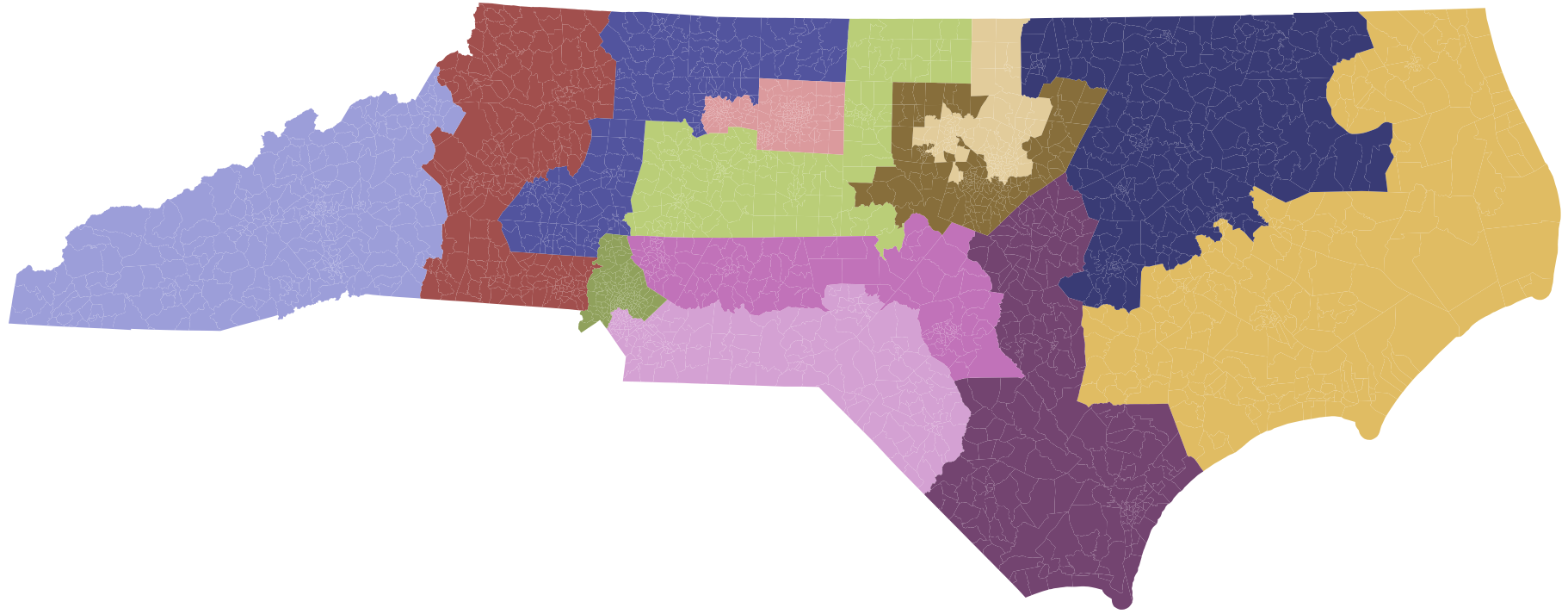}
    \caption{2020 NC congressional districting map after one Recom operation.}\label{subfig:recom}
\end{subfigure}
\caption{Here, the Recombination algorithm is used to generate a new districting plan starting from the 2020 enacted plan in North Carolina.}
\Description[Recom]{NC congressional districting maps after one Recom step}
\end{figure}

An example of Recom after one step performed on the 2020 Congressional Districting Plan for North Carolina is shown in Figure ~\ref{subfig:recom} where two new districts emerge to create a slightly different plan than the original map. Unlike the Flip node algorithm, which can take millions of steps until a new plan is achieved that is compact and differs from the original plan due to the small movements between each state, Recom can significantly alter a given plan in a short number of steps. Figure ~\ref{fig:biased_maps} shows two different maps obtained by running a Markov chain for 100 steps using techniques referred to as hill climbing (Figure ~\ref{subfig:hill_100}) and short bursts (Figure ~\ref{subfig:shortburst_100}) starting from the 2020 Congressional Districting Plan for North Carolina. 

\label{fig:nc-recom}

\subsection{Partisan Metrics}\label{subsec:partisan}

Given a districting plan, it is of interest to be able to determine quantitatively if it has been gerrymandered. There exist large families of metrics that can be associated to a particular plan we refer to as \emph{partisan metrics}. Most existing metrics utilize notions of proportionality or symmetry to quantify the fairness of a plan. In Figure \ref{fig:ohio-map-matters}, the map proposed by the Ohio Citizens' Redistricting Commission had proportionality in mind, where Republicans had around $\sim$54\% of the statewide voteshare and thus had 8 Republican leaning congressional districts, representing $\sim$54\% of the 15 possible congressional seats. Partisan symmetry on the other hand, captures the idea that if Democrats in North Carolina win 60\% of the seats with 52\% of the votes, then if Republicans won 52\% of the votes, we would also expect them to win around 60\% of the seats. If we let $S(V)$ be the seat share for a party given a vote share $V$, partisan symmetry can be expressed by the equation: $S(V) = (1-S(1-V))$.

Here we consider the following partisan metrics: partisan bias, partisan gini, mean-median, efficiency gap, and safe seats. Each can be viewed as a function $\omega_{*}:\Sigma \to \mathbb{R}$ which will assign to some map a signed score indicating its partisan bias. 

\subsubsection{Overview of select partisan metrics}

The \emph{mean-median} score is a measure that quantifies how far a party is from achieving half of the seat share in their state after receiving half of the votes. Given a vote share vector $(v_1,\hdots,v_k)$ where each $v_i$ represents the Republican (or Democratic) vote share in district $i$, we compute the mean-median score by subtracting the average vote share of the $v_i$ from the median vote share as follows:

\[
\omega_{MM}=\text{median}\{v_i\}-\text{mean}\{v_i\}
\]

Note that the mean-median score will also give a notion of the skew of the vote share distribution for a given election. When the mean and median vote share are close, the distribution of votes is more symmetric and in line with the idea of partisan symmetry being a measure of fairness. A large mean median score implies that a distribution is skewed in favor of one party over another. An ideal mean-median score is 0.

The \emph{partisan bias} of a plan quantifies how many more than half of the seats they will gain by having half of the votes. It tells on average, how much a plan deviates from the partisan symmetry standard for a given vote share in an election. Unlike the mean-median score, which is votes dominated, partisan symmetry is seats dominated. It is given by the formula below:

\[ 
\omega_{PB}=\frac{S(V)-(1-S(1-V))}{2}.
\]

The \emph{partisan gini} computes the average value of the partisan bias over the interval $[0,1]$ versus the partisan bias which can be computed for each possible vote share $V \in [0,1]$. Note that sometimes the interval is restricted to avoid extremes of the vote distributions. We choose the entire interval as we are interested in generating plans with extreme bias. The partisian gini score can be computed using the following formula:

\[ 
\omega_{PG}=\frac{1}{2}\int_0^1\left|S(V)-(1-S(1-V))\right|dV
\]

The \emph{efficiency gap}, introduced in \cite{stephanopoulos2015partisan}, provides a measure of how well a given party converts votes to seats by quantifying the number of wasted votes for a given election into a signed score. A wasted votes is any vote for the losing party in a given district, or any vote for the winning party over the 50 percent vote share needed to win. For consistency, we provide a formulation of the efficiency gap solely dependent on a given party's vote share $V$ and corresponding seat share $S(V)$:

\[
\omega_{EG}=\left(S(V)-\frac{1}{2}\right)-2\left(V-\frac{1}{2}\right)
\]

Lastly, a \emph{safe seat} in an election is one that is considered to be very secure for a particular party, that is, given the distribution of votes, it is highly unlikely that this seat will go to the opposite party. A competitive district, where the voting distributions between each party are similar with outcomes highly susceptible to things such as campaigning, voter turnout, etc. will produce a marginal seat. Thus, one way a party can maximize representation is to have a districting plan that maximizes their number of safe seats.

\section{Biased Ensembles}

We empirically assess the power of this statistical test from Theorem ~\ref{thm:eatest} in two steps--(1) first we generate a large collection of maps intentionally biased with respect to different partisan metrics, and (2) then we use these maps to probe the updated statistical test introduced in \cite{chikina2020separating}. In this section, we outline our methodology for creating such a collection (step 1) and analyze our results. Our code is available at \href{https://github.com/HarlinLee/gerrypowers}{https://github.com/HarlinLee/gerrypowers}. North Carolina shapefiles and presidential election data were downloaded from {\color{blue} \url{https://github.com/mggg-states/NC-shapefiles}}, and the code that generates biased ensembles is built on top of the Python GerryChain package \cite{gerrychain}.

\subsection{Generating Biased Ensembles}\label{subsec:generatingbiased}

There are two prevailing methods for generating collections of redistricting plans, referred to as ensembles, that are biased with respect to certain partisan metrics. The first method is known as `hill climbing,' introduced in \cite{duchin2022homological} to generate ensembles biased towards increasing the number of safe seats for a particular party. The second method, called `short bursts,' was introduced in \cite{cannon2023voting} to optimize plans that maximize the number of majority-minority districts in a given state.

The hill climbing method utilizes a Metropolis-style weighting that prioritizes the acceptance of maps within the proposal step of the Markov Chain that are similar or more extreme than the current plan with respect to a particular metric. To do this we first choose a political party $P \in \mathcal{P}$ and a partisan metric of interest $\omega_P: \Sigma \to \mathbb{R}$ (e.g. efficiency gap) such that $\omega_P(X') \ge \omega_P(X)$ if a map $X' \in \Sigma$ is more biased in favor of party $P$ than another map $X \in \Sigma$. If $X$ is our current map and $X'$ is the proposed map, we accept $X'$ with probability $e^{\beta (\omega_P(X') - \omega_P(X))}$ for some $\beta > 0$. Note that when $\omega_P(X') \ge \omega_P(X)$, the new map $X'$ is always accepted. When $\omega_P(X') < \omega_P(X)$, the new map is accepted with some probability to prevent that chain from getting stuck in local optima. This is summarized in Algorithm~\ref{alg:hill}.

\begin{algorithm}
\caption{Generating biased ensembles using Hill Climbing \cite{duchin2022homological}}
\label{alg:hill}
\begin{algorithmic}[1]
\Procedure{HillClimbing}{Map $X$, Label function $\omega$, Steps $N$, Temperature $\beta$}
    \State Initialize $X_0 \gets X$.
    \For {$i= 1, \ldots, N$}
        \State $X_i \gets$  Recom($X_{i-1}$).
        \State $\delta \gets \omega(X_{i}) - \omega(X_{i-1})$.
        \State Generate uniform random number $p \in [0, 1]$.
        \If{$p \ge \exp(\beta\delta)$}
        \State Reject the candidate map, i.e. $X_i \gets X_{i-1}$.
        \EndIf
        \State Save map $X_i$.
	\EndFor
	\State Return all saved maps.
\EndProcedure
\end{algorithmic}
\end{algorithm}

Short bursts are a type of Markov Chain that performs a unbiased random walk for a small number of steps, referred to as a `burst.' After each burst a metric is computed from each state in the Markov Chain, and then whichever state has the highest value becomes the initial state for the next short burst. The number of steps is referred to as the burst length. Unlike hill climbing, where chains are often run for tens and hundreds of thousands of steps, in short bursts the burst length could be as small as 5 steps. In \cite{cannon2023voting} they found optimal results for with a burst length equal to 10. This algorithm is described in Algorithm~\ref{alg:shortburst}.

\begin{algorithm}
\caption{Generating biased ensembles using Short Bursts \cite{cannon2023voting}}
\label{alg:shortburst}
\begin{algorithmic}[1]
\Procedure{ShortBurst}{Map $X$, Label function $\omega$, Burst length $k$, Runs $N$}
\State Initialize $X_0 \gets X$.
    \For {$i= 1, \ldots, N$}
        \For {$j = 1, \ldots, k$}
            \State $X_j \gets$  Recom($X_{j-1}$).
            \State Save map $X_j$.
    	\EndFor
     \State Set $X_0 \gets \arg\max_{X_i}~\{\omega(X_1), \omega(X_2), \ldots, \omega(X_k)\}$.
    \EndFor
	\State Return all saved maps.
\EndProcedure
\end{algorithmic}
\end{algorithm}

\begin{figure}[htp]

\begin{subfigure}{\linewidth}
\centering
    \includegraphics[width=0.75\linewidth]{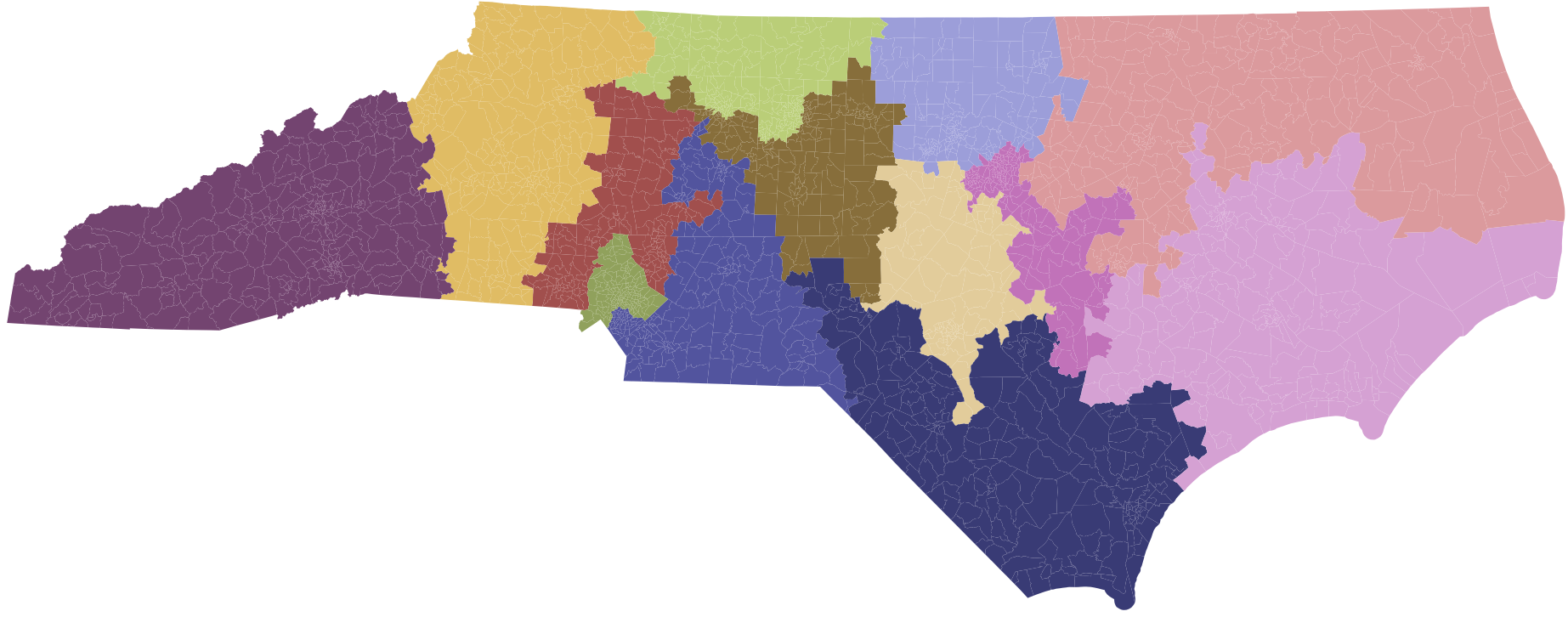}
    \caption{NC congressional district map after hill climbing (Alg.~\ref{alg:hill}) with $N=100$.}\label{subfig:hill_100}
\end{subfigure}
\begin{subfigure}{\linewidth}
\centering
    \includegraphics[width=0.75\linewidth]{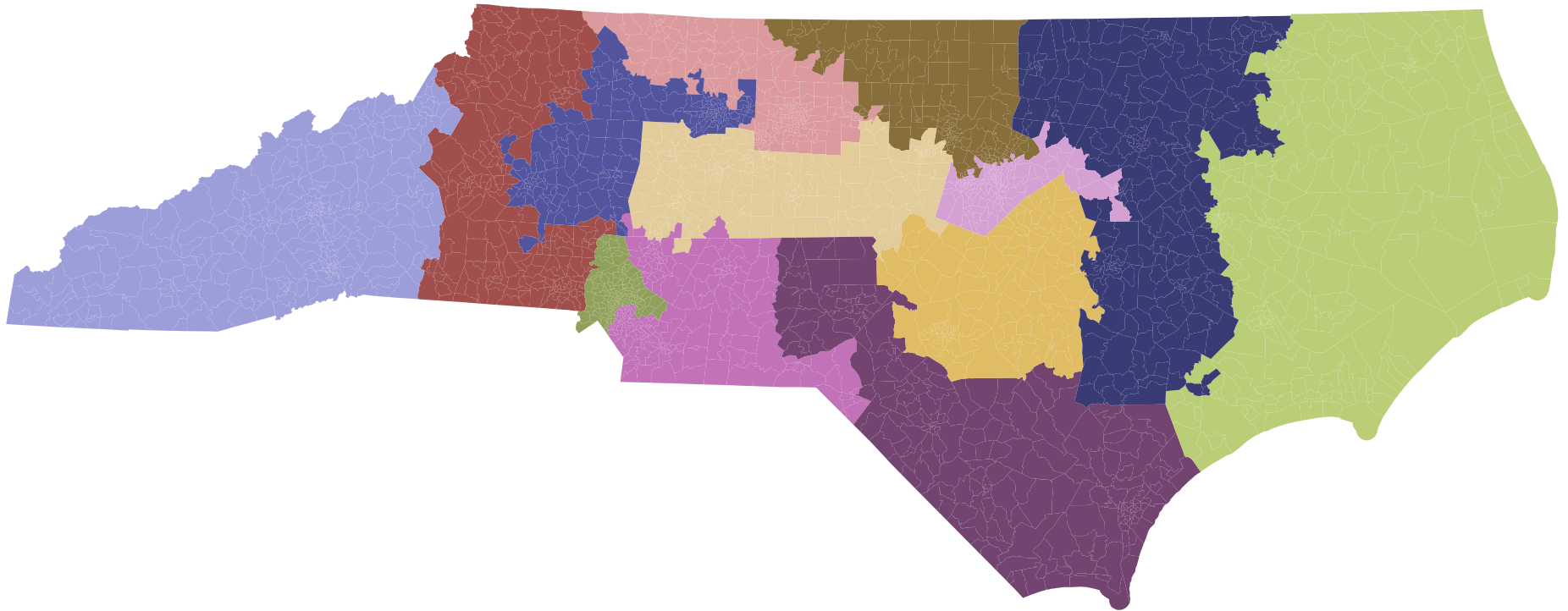}
\caption{NC congressional district map after short burst (Alg.~\ref{alg:shortburst}) with $N=20, k=5$.}\label{subfig:shortburst_100}
\end{subfigure}
\caption{Examples of maps from biased chains. Biasing was done to favor the Republican party according to the mean median gap using 2016 Presidential election data. Coloring is arbitrary.}\label{fig:biased_maps}
\Description[Biased maps]{Biased map examples}
\end{figure}

We consider how each of these methods behave in generated ensembles that are biased with respect to several partisan metrics of interest--efficiency gap, mean-median, partisan bias, partisan gini, and safe seats. Note that there are many other partisan metrics of interest amenable to this approach including declination, GEOmetric \cite{campisi2022geography}, mean thirdian \cite{gerrychain}, ordered marginal medians, smooth seat count deviation, representative index, and gerrymandering index \cite{herschlag2020quantifying}. Our focus is on those metrics that have commonly emerged in redistricting litigation. We used the Recom algorithm via short bursts for 50,000 steps (10,000 runs with burst length of 5) and hill climbing method for 50,000 steps. Figure ~\ref{fig:nc-recom} shows maps after 1 Recom operation (Fig.~\ref{subfig:recom}), while Figure ~\ref{fig:biased_maps} shows maps 100 steps of Hill Climbing (Fig.~\ref{subfig:hill_100}), and 20 runs of Short Burst with 5 burst length (Fig.~\ref{subfig:shortburst_100}).

\begin{figure}[htbp]
    \centering
    \begin{subfigure}{0.35\linewidth}
        \centering
        \includegraphics[width=\linewidth]{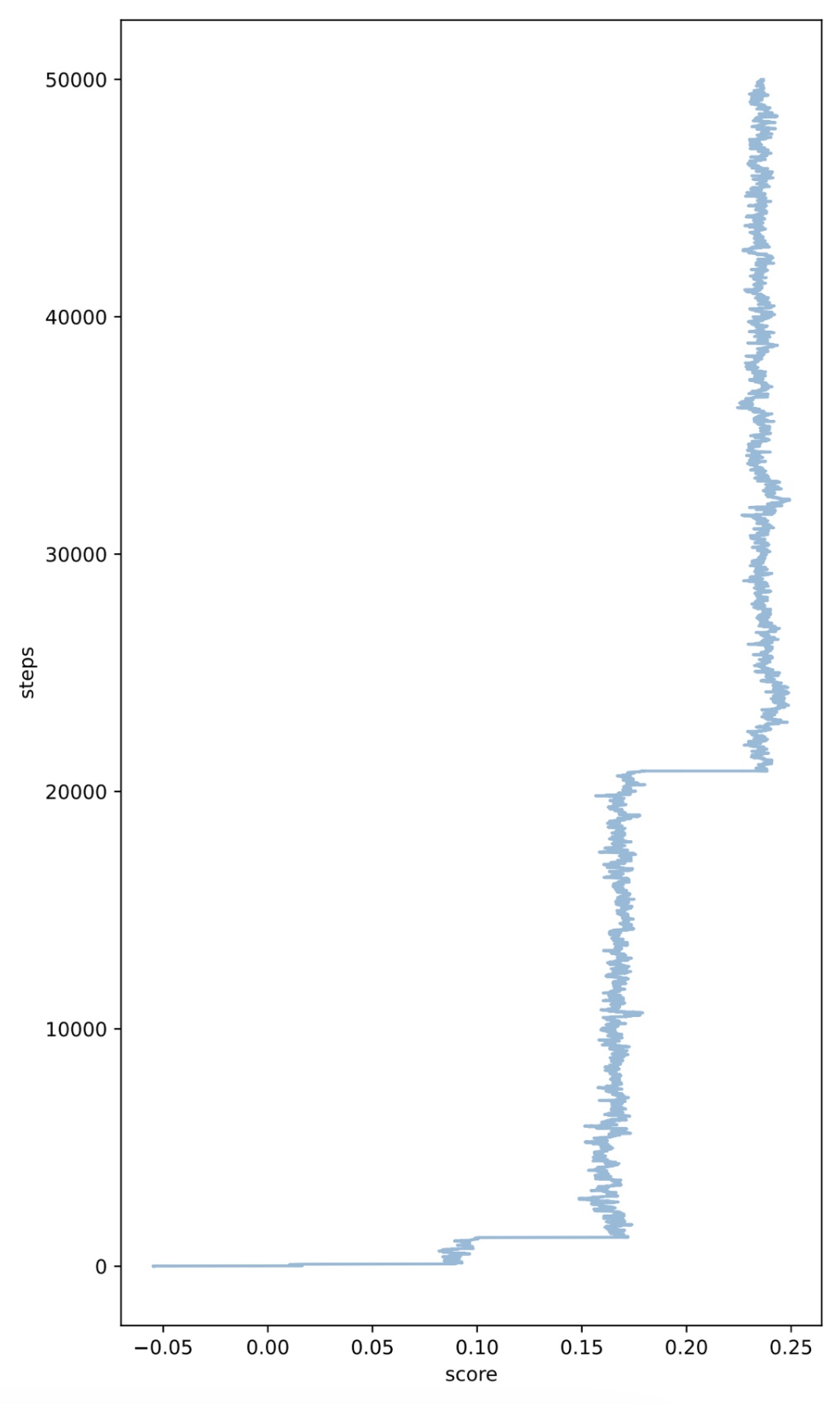}
        \caption{Hill Climbing (D), Efficiency Gap}
        \label{fig:fig1}
    \end{subfigure}%
    \begin{subfigure}{0.35\linewidth}
        \centering
        \includegraphics[width=\linewidth]{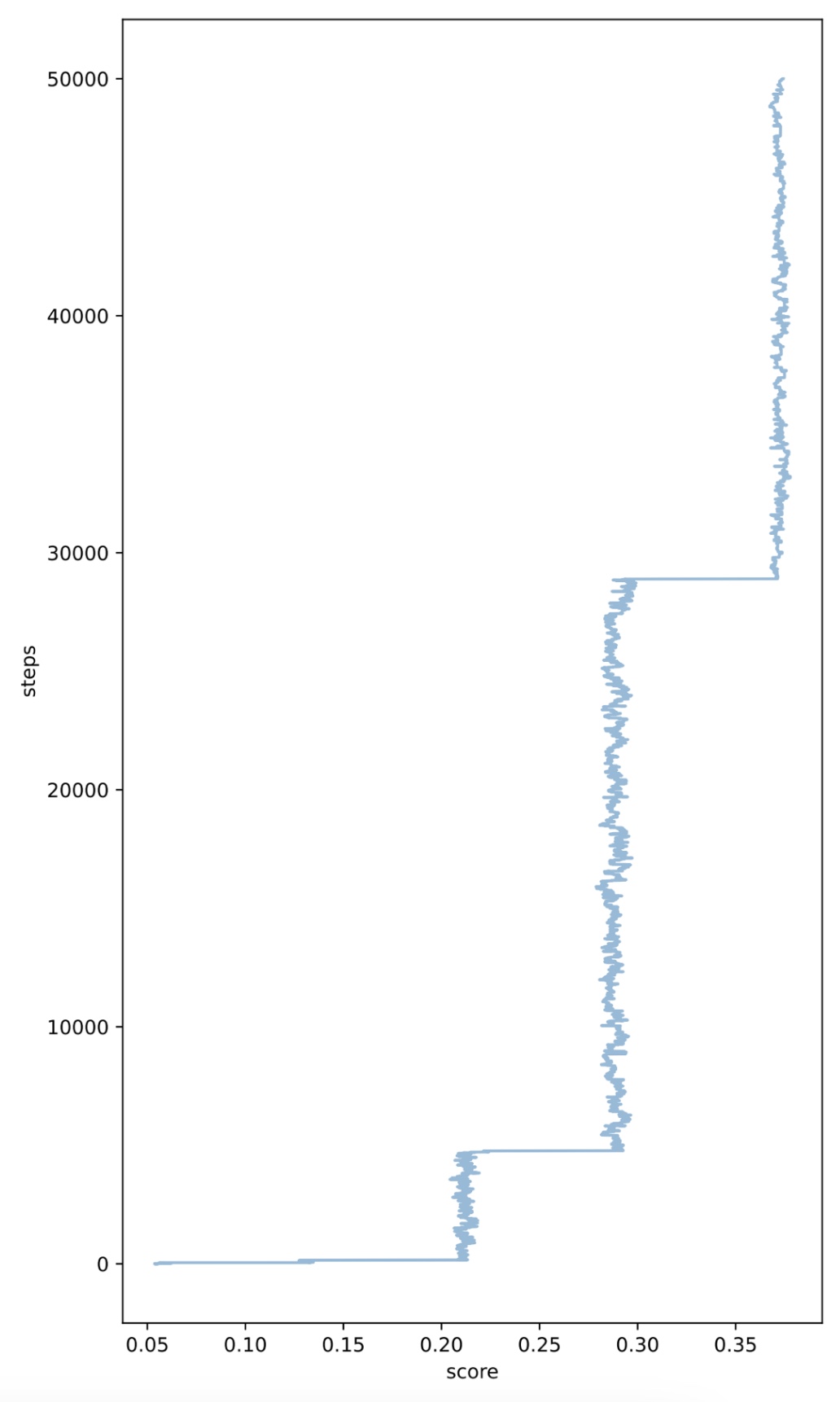}
        \caption{Hill Climbing (R), Efficiency Gap}
        \label{fig:fig2}
    \end{subfigure}
    \vfill
    \begin{subfigure}{0.35\linewidth}
        \centering
        \includegraphics[width=\linewidth]{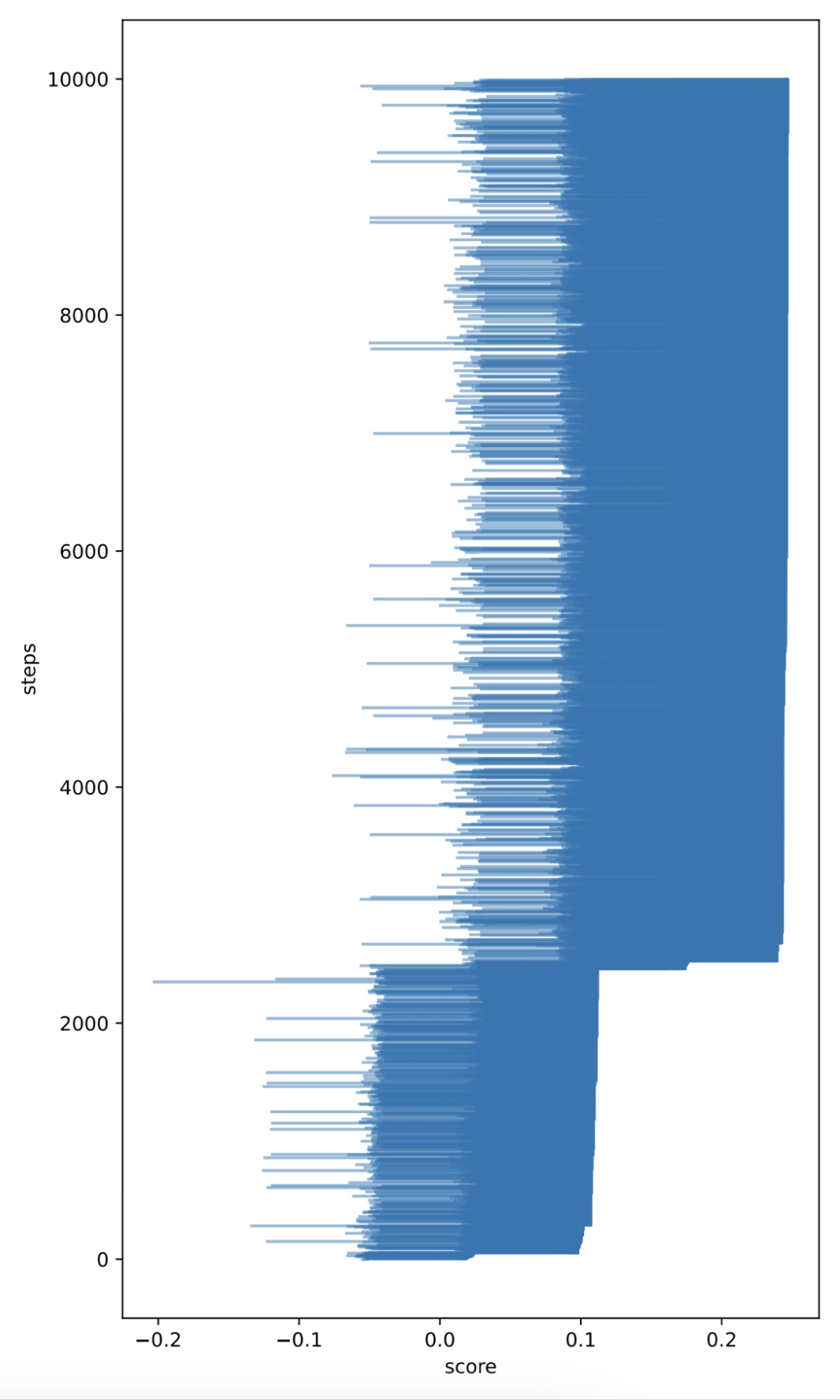}
        \caption{Short Bursts (D), Efficiency Gap}
        \label{fig:fig3}
    \end{subfigure}%
    \begin{subfigure}{0.35\linewidth}
        \centering
        \includegraphics[width=\linewidth]{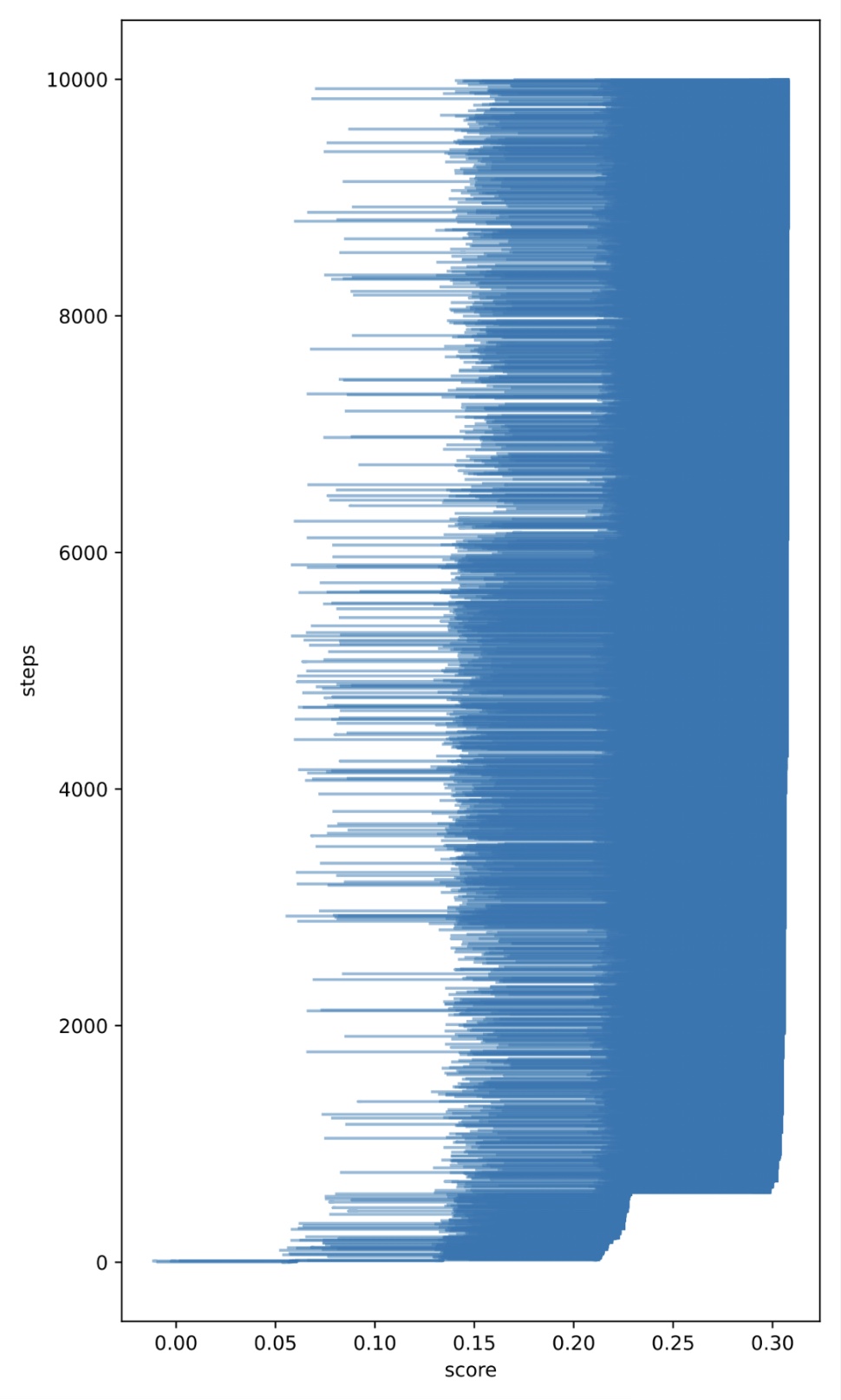}
        \caption{Short Bursts (R), Efficiency Gap}
        \label{fig:fig4}
    \end{subfigure}
    \caption{Values of efficiency gaps in ensembles biased for Republicans and Democratics using hill climbing and short bursts respectively after 50000 steps for the 2016 Presidential Election in North Carolina.}
    \label{fig:eg-chains}
    \Description[Biased chain map values]{Efficiency gap plot for biased chain maps}

\end{figure}

\subsection{Analyzing Biased Ensembles}

The methodology described in Section ~\ref{subsec:generatingbiased} was successful in producing plans with extreme values for our chosen metrics. For example, in Figure ~\ref{fig:eg-chains} we observe how the efficiency gap of the redistricting plans is changing as the number of steps in the biased random walk increases. The results in maximum efficiency gap achieved are similar between the chains utilizing hill climbing and those utilizing short bursts for both parties, but the extreme scores are achieved much earlier with short bursts--the maximum efficiency gap for random walks biased in favor of the Democrats is around 0.25 in each chain, however this value is achieved in around 11,000 steps $(2100\cdot 5)$ for short bursts and 21000 steps for hill climbing. Note that the biased ensembles are indeed pushing plans to extreme scores for this particular partisan metric. In each of the subfigures in Figure~\ref{fig:eg-chains}, the efficiency gap equals or exceeds 0.25, whereas \cite{stephanopoulos2015partisan} suggested that a plan be flagged as a gerrymander if the efficiency gap exceeded $|0.08|$.

\begin{figure}
    \centering
    \includegraphics[trim={0 0 0 1cm}, clip,width=.5\textwidth]{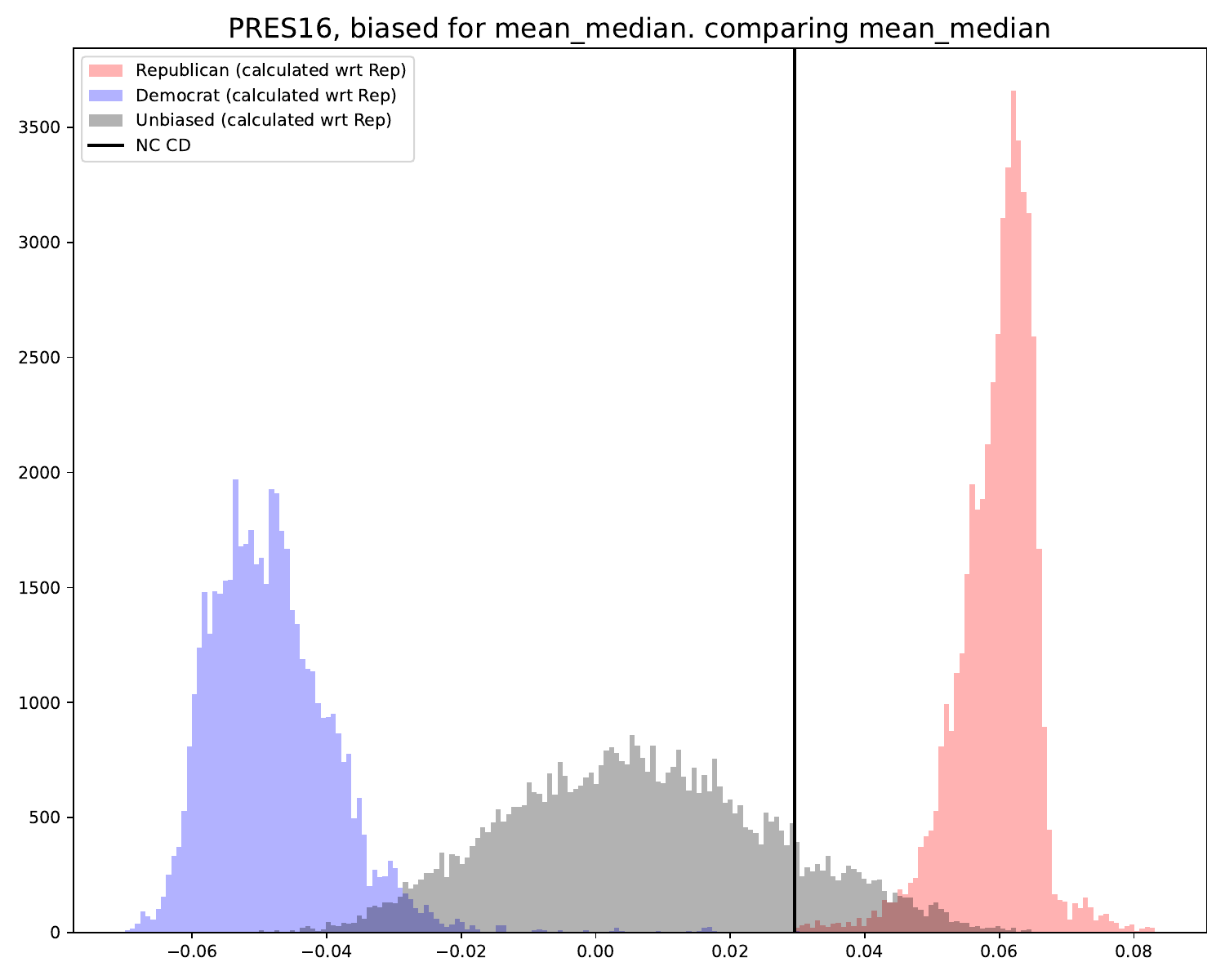}%
    \includegraphics[trim={0 0 0 1cm}, clip,width=.5\textwidth]{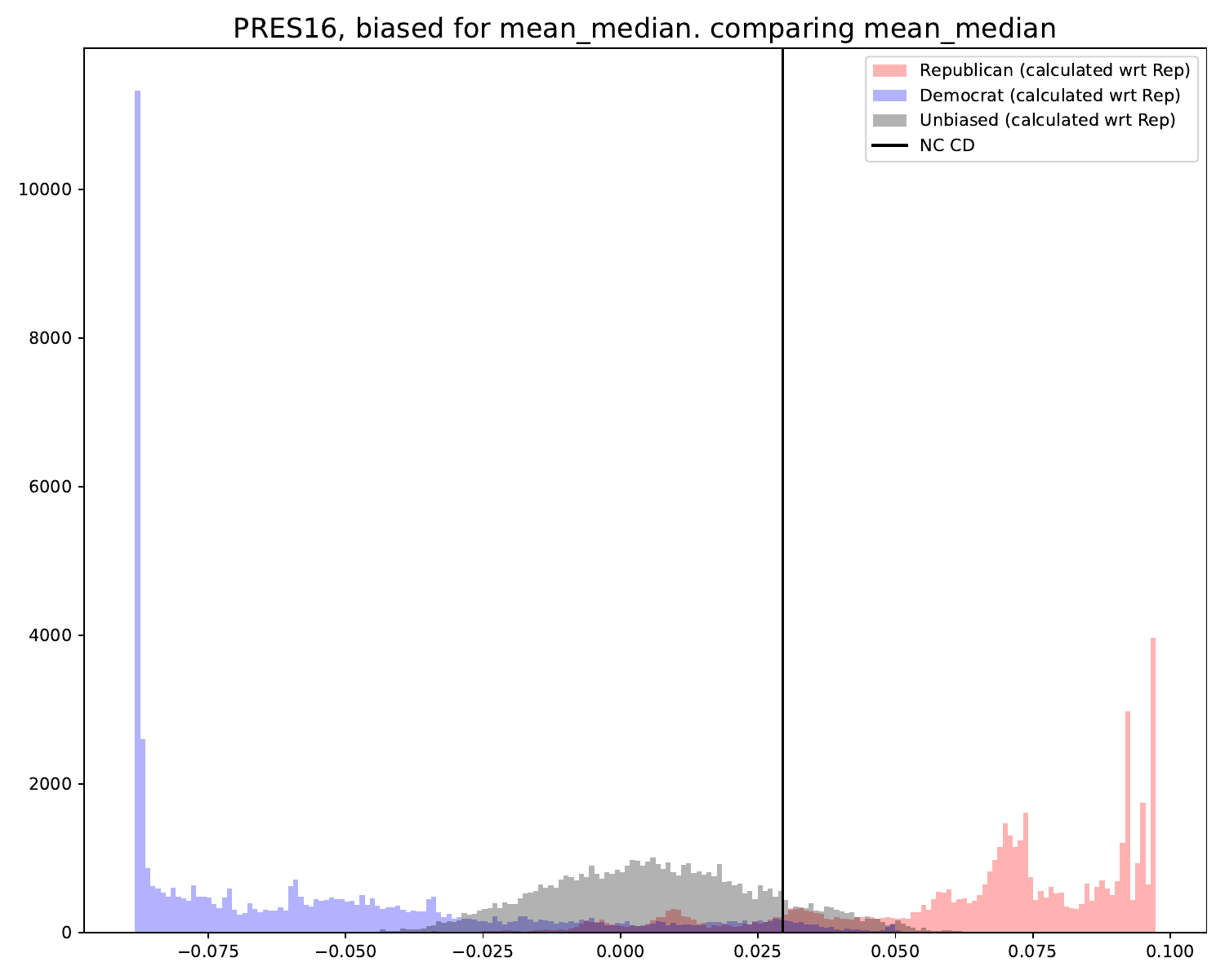} 
    \caption{Histograms of mean median gap values from biased NC maps. Biased chains are run to maximize mean median gap with hill-climbing (left, $N=50000$) and short burst (right, $N=10000, k=5$) using 2016 Presidential election data. Chains start at the map in Fig.~\ref{fig:original_map}, whose mean median gap value is marked in solid vertical line. Blue is distribution of maps biased for Democrats, and red is for Republicans. Gray is for neutral Recom chain of length 50000 and are the same in both subplots.}
    \label{fig:biased_mean_median}
    \Description[Histograms]{Histograms of biased maps}
\end{figure}

The phenomenon observed in Figure ~\ref{fig:eg-chains} was indicative of a general pattern where the extreme values of the partisan metrics were in similar ranges for both the hill climbing and short burst method. This pattern held over both the 2012 and 2016 Presidential elections in North Carolina for ensembles biased for mean-median, efficiency gap, and safe seats. In the 2016 Presidential election in North Carolina the safe seats were 11R to 10R, 8D to 7D for Republican and Democratic favoring chains for hill climbing and short bursts respectively. The mean-median scores were 0.37R to 0.3R, 0.25D to 0.25D for Republican and Democratic favoring chains for hill climbing and short bursts respectively. Recall from Section ~\ref{subsec:partisan} that an ideal mean-median score is 0 and that in 2012 and 2016 there were 13 congressional districts, and thus seats, in North Carolina. Figure ~\ref{fig:biased_mean_median} show histograms that highlight the range and frequency of mean-median scores in the two types of biased ensembles for the two elections above compared to a neutral ensemble, that is one generated via hill climbing without any intentional bias.

There were discrepancies between the hill climbing and short bursts methods for partisan gini and partisan bias. For the 2012 North Carolina presidential election the partisan bias scores were similar for Democratic favoring chains between the two methods, but the hill climbing method produced slightly more extreme partisan bias scores for the Republican favoring chain at 0.35 versus 0.275 for the Republican favoring chain via short bursts. For partisan gini in this election, the Republican favoring chain for hill climbing produced extreme scores around 0.06 while the short burst method produced extreme scores around 0.12. An ideal partisan gini score would be closer to 0, with higher values indicating more disproportionate outcomes between parties across the districts. Meanwhile a partisan bias score between 0.7 and 0.10 is an indicator of bias, and so the chains from both methods are produced maps with obvious bias with respect to that metric.

We also observed how biased an ensemble with respect to one metric effected the scores with regards to other metrics. Figure ~\ref{fig:Interactionplots} shows a scatterplot of interactions between our chosen metrics. 

\begin{figure}
\centering
   \includegraphics[trim={0 0 0 1cm}, clip, width=\linewidth]{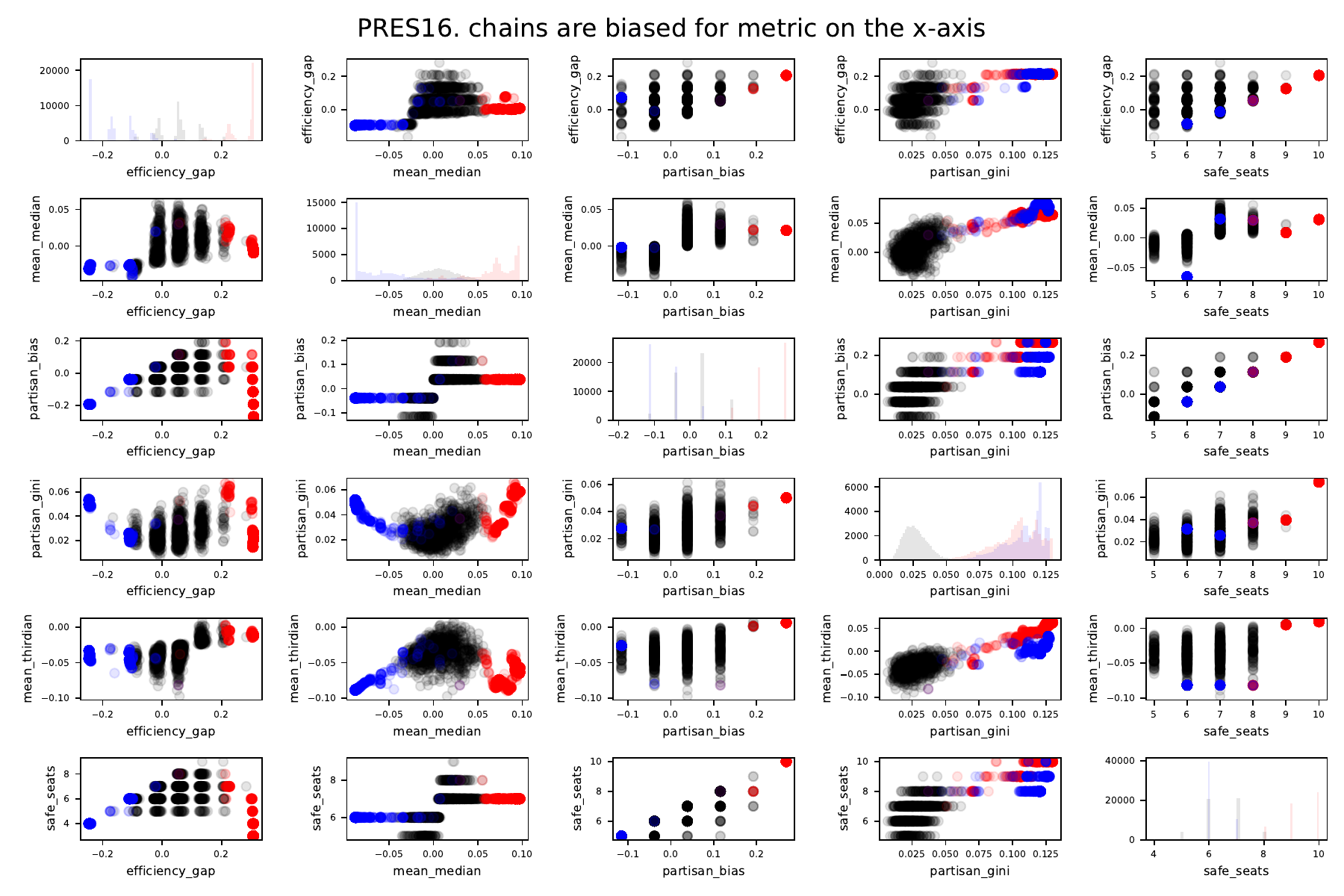}
    \caption{Scatter plots indicate nonlinear relationships between bias metrics. Histograms of mean median gap values from biased NC maps. Blue points are maps biased for Democrats, and red is for Republicans. Gray is for neutral Recom chain. In each subplot, horizontal axis is the metric the chain was biased towards, and vertical axis is the metric we are measuring. When the metric used to bias the chains and the metric measured from the resulting chains are the same, we plot the histograms as in Fig.~\ref{fig:biased_mean_median}. Biased chains were generated via short burst (Alg.~\ref{alg:shortburst}) for $N=10000, k=5$ using 2016 presidential election data. For readability, every 50th map in the chain was sampled for the scatter plots. }\label{fig:shortburst_rep_scatter}
    \Description[Scatter plots]{Scatter plots of biased maps between bias metrics.}
\end{figure}

\section{Empirical results}\label{sec:EmpiricalResults}
To test the effectiveness of this test, we empirically evaluate the two types of error important to hypothesis tests: Type I ($\alpha$) and Type II ($\beta$). The power (1-$\beta$) and Type I error are assessed across (i) three different variable settings: election year $y$, gerrymandering metrics (label function $\omega$), political party $P$ and (ii) two different parameter settings: effect size $\varepsilon \text{ and sample size } k$.

\subsection{Type I Error}
Type I error, often referred to as the false positive rate, is predefined before the test as $\alpha$. It signifies the probability of incorrectly rejecting the null hypothesis when the state $\sigma^0$ is not an $(\alpha, \varepsilon)-$outlier. Empirically testing the Type I error rate includes setting $\alpha$ ($\alpha=0.05$ is used for Section~\ref{sec:EmpiricalResults}), conducting multiple hypothesis tests under the null distribution, and calculating the percentage of times the null hypothesis was rejected, or $\hat \alpha$. The unbiased chains $\Sigma^{0}$ serve as a surrogate for the null distribution.

For a given variable/parameter combination ($y,\omega,P,\varepsilon,m,k$), Type I error was assessed by sampling 100 maps $\{\sigma_1^0, \ldots, \sigma_{100}^0 \} \in \Sigma^0$ at random and calculating a p-value using Equation~\eqref{eq:NullDistribution} to test if each map is an $(\alpha, \varepsilon)-$outlier or not. In general, the Type I error rate was zero for all variable combinations. The only exception was when the label $\omega$ was efficiency gap, the Type I error rate was $\hat \alpha=0.02$. This result is consistent with the test in Theorem \ref{thm:eatest}, which indicates that the Type I error for every unbiased chain $\Sigma^0$ should be at most $\alpha$. We calculated $\hat \alpha$ as a validation measure for using $\Sigma^0$ and $\Sigma^A$ as stand-ins for the null distribution and the alternative distribution at various variable/parameter combinations, where $\Sigma^A$ denotes a biased chain.

\subsection{Power Analysis}

\begin{algorithm}[ht]
\caption{Implementation of $(\alpha, \epsilon)$-outlier test in \cite[Theorem 3.1]{chikina2020separating}.}
\label{alg:outlier}
\begin{algorithmic}[1]
\Procedure{OutlierTest}{Map $X$, Label function $\omega$, Significance level $\alpha$, Outlier level $\epsilon$, Number of trajectories $m$, Steps $k$}
    \State Initialize $\rho \gets 0$.
    \For {traj $ = 1,\ldots, m$}
    \State Initialize $X_0 \gets X$.
    \For {$i=1,\ldots,k$}
        \State $X_i \gets $flipnode($X_{i-1}$).
 \EndFor
 \State Set $c$ as the number of $X_i$ in $i=1, \ldots, k$ such that $\omega(X_i) < \omega(X)$.
 \If{$c > (1-\epsilon)k$}
 \State Increase $\rho \gets \rho + 1$ since $X$ is an upper $\epsilon$-outlier in this trajectory.
 \EndIf
 \EndFor
 \State Calculate $p \gets \exp(-\min (r^2\sqrt{\frac{\alpha}{2\epsilon}}/3m, r/3))$, where $r = \rho - m\sqrt{\frac{2\epsilon}{\alpha}}$.
 \State Return $p\le \alpha$, i.e. whether $\omega(X)$ is an $(\alpha, \epsilon)$-outlier.
\EndProcedure
\end{algorithmic}
\end{algorithm}

\begin{algorithm}[ht]
\caption{Power calculation.}
\label{alg:power}
\begin{algorithmic}[1]
\Procedure{PowerAnalysis}{Biased ensemble $\{X_i\}_{i=0}^N$, Number of samples $n$, Outlier test parameters $\omega,\alpha, \epsilon, m, k$}
    \State Uniformly sample $n$ maps from the MCMC chain $\{X_i\}_{i=0}^N$.
    \State Initialize $c \gets 0$.
    \For {each sampled map $X_i$}
    \State detected $\gets $ OutlierTest($X_i, \omega, \alpha, \epsilon, m, k$).
    \If{detected}
    \State Increase $c \gets c + 1$.
    \EndIf
 \EndFor
	\State Return $c$.
\EndProcedure
\end{algorithmic}
\end{algorithm}

The power of a statistical test, in our context, is the likelihood that the test in Theorem \ref{thm:eatest} can detect a gerrymandered map; this is the true negative rate, or $(1-\beta)$ where $\beta$ is false negative rate. In order to calculate the power of this test empirically, the biased chains $\Sigma^A$ serve as surrogate distributions for the distribution of gerrymandered maps. Empirically testing the power includes setting $\alpha$, conducting multiple hypothesis tests under the alternative distribution, and calculating the percentage of times the null hypothesis was rejected. 

For a given variable/parameter combination ($y,\omega,P,\varepsilon,m,k$), power was assessed by sampling 100 maps $\{\sigma^0_1, \ldots, \sigma^0_{100} \} \in \Sigma^A$ at random and calculating a p-value using Equation~\eqref{eq:NullDistribution} to test if each map is an $(\alpha, \varepsilon)-$outlier or not. To perform this hypothesis test $m$ reversible Markov chains are run with $k$ steps each. 

As stated in Section~\ref{sec:StatisticalTest}, power relies on $\alpha, \varepsilon,m$ and $k$, where higher values of these parameters will lead to more powerful tests. However, we also want $\alpha$ and $\varepsilon$ values which are meaningful (i.e. help to quantify the unlikeliness of a state) and $m$ and $k$ values which limit computational cost (i.e. low sample sizes). We are also interested in if the power of the test changes for different election variables and if there are variable interactions. First we focus on how the parameters values effect the test and then focus on election variables.

\subsubsection{Parameters: $k$ and $\varepsilon$}
\begin{figure}[htp]
\centering
    \includegraphics[width=0.7\linewidth]{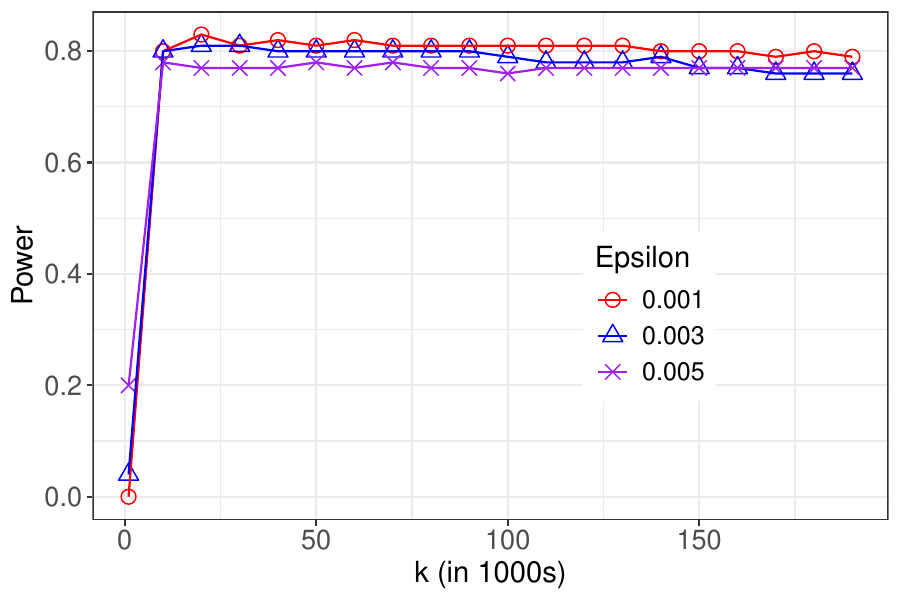}
    \caption{Plot of Power of statistical test (y-axis) as the number of steps $k$ in the chain get larger (x-axis) for different values of $\varepsilon$ (color/shape). }\label{fig:PowerK}
    \Description[line plots]{line plots that show power saturates as $k$ increases.}
\end{figure}

Larger sample sizes ($k$ and $m$) and larger effect sizes to find significance ($\varepsilon$) are conventionally thought to lead to higher power. On the other hand, smaller $k$ leads to less computational cost in general, and smaller $\varepsilon$ values may be needed for real world problems. In this subsection, we investigate how changes in $k$ and $\varepsilon$ lead to changes in the power of a particular $\Sigma^A$ (efficiency gap, democratic, year 2012). For the same $\Sigma^A$, we tested power for different values of $k$ and $\varepsilon$ where $k=\{200000, 190000, 180000, \ldots, 20000, 10000, 1000 \}$ and $\varepsilon=\{0.001, 0.003, 0.005 \}$. There did not appear to be an interaction between $\varepsilon$ and $k$, as seen in Figure~\ref{fig:PowerK}, and the power did not change for $k$ values until $k$ became very small (e.g. $1000$ steps). We use an asymptotic regression model to help describe the limit growth of Power as $k$ goes to infinity as described below:

\begin{align}\label{eq:AsymReg}
    \text{Power}(k) = a-(a-b) \exp(-ck)
\end{align}

where $a$ is the maximal attainable power, $b$ is the y-intercept, and $c$ is proportional to the relative rate of Power increase while $k$ increases. Using equation~\eqref{eq:AsymReg}, with a re-scaled $k/1000$ for model convergence, the model estimated that the maximal attainable power of approximately 0.79 for this $\Sigma^A$ chain, and this occurs around $k=247.68$ (after re-scaling), where the power is near its maximal value. In other words, if we let our chains only go approximately $248$ steps, this will lead to the same power as a chain which goes $200000$ steps. The results are shown in Figure~\ref{fig:PowerK}, where we can see that maximal power is attained at low $k$ values across all $\varepsilon$ values.

\begin{figure}[htp]
\centering
    \includegraphics[width=0.8\linewidth]{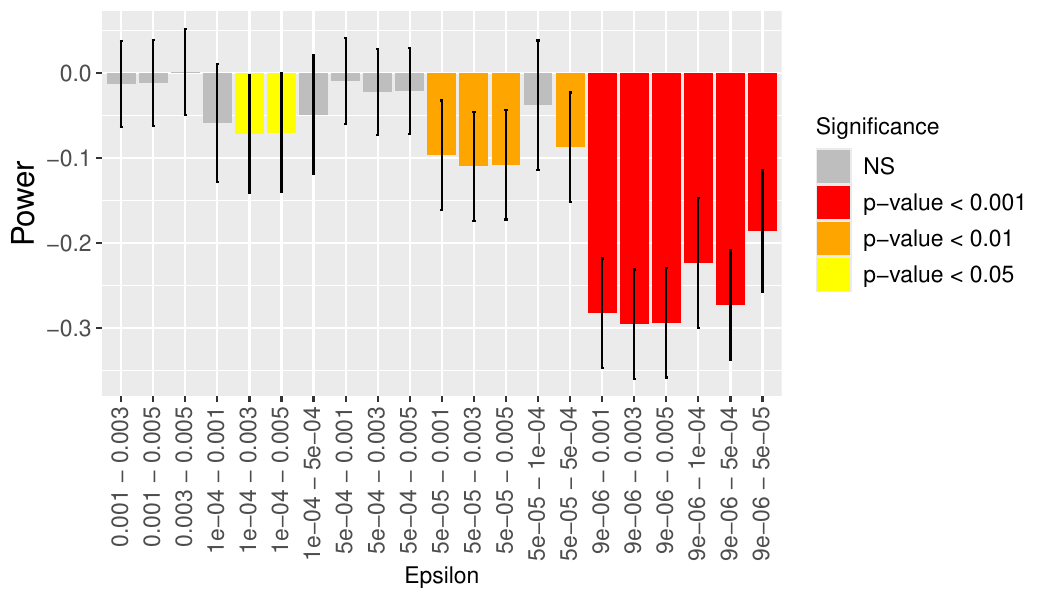}
    \caption{Tukey HSD results for different values of $\varepsilon=\{ 0.001, 0.003, 0.005, 0.0005, 0.0001, 0.00005, 0.000009\}$. The color indicates the statistical significance of the p-value for each pairwise test (x-axis). The difference in the mean power between $\varepsilon$ values is shown in the $y-axis$.}\label{subfig:HSDEpsilon}
    \Description[HSD plots]{power decreases with lower epsilon.}
\end{figure}

Looking at Equation~\ref{eq:NullDistribution}, the effect size must have values within the range of $1/k < \varepsilon < \alpha/2$ in order to have non-trivial results. For this section, we tested $\varepsilon$ values of \sloppy$\{0.001, 0.003, 0.005, 0.0005, 0.0001, 0.00005, 0.000009\}$ on different chains $\Sigma^A$ to see if $\varepsilon$ values have a significant effect on the power of the test. The results of this simulation are shown in Figure~\ref{subfig:HSDEpsilon}. The HSD pairwise tests between $\varepsilon$ values indicate that smaller values of $\varepsilon$ lead to much less power (in many cases the power is close to $0$). However, larger $\varepsilon$ lead to a higher power and there is not a significant different between $\varepsilon=\{0.001, 0.003, 0.005 \}$. Since higher $\varepsilon$ values lead to a higher power in the test, we chose $\varepsilon=0.001$ for the rest of the section.

Next we looked at how the power changed depending on election variables such as (1) metric the chain $\Sigma^A$ was biased for, (2) election year, and (3) party the chain $\Sigma^A$ was biased for. We first focus on each variable individually and calculate the power when $\varepsilon=0.001, \alpha=0.05, m=32$ and $k=200000$ as these parameter values lead to the highest power. We focus on the main effects since there was no evidence of an interaction between these variables.

\subsubsection{Metrics: $\omega=\{$mean median, efficiency gap, partisan bias, partisan gini, and safe seats$\}$} 

The partisan metrics are described in detail in Section ~\ref{subsec:partisan}. Each metric in $\omega$ measures different aspects of fairness in an election. In the table below, we report the average power for each metrics independent of all other variables where the test for each $\omega$ is run on biased chains $\Sigma^A$ which are biased for that $\omega$ value.  

\begin{center}
\begin{tabular}{c c c c c} 
 \toprule
 \textbf{Efficiency gap} & \textbf{Mean median} & \textbf{Partisan bias} & \textbf{Partisan gini} & \textbf{Safe seats}\\ \midrule
 $0.76 (\pm 0.009)$ & $0.62 (\pm 0.01)$ & $0.005 (\pm 0.001)$ &  $0.98 (\pm 0.001)$ & $0.1 (\pm 0.02)$ \\ \bottomrule 
\end{tabular}
\end{center}

\begin{figure}[H]
\centering
\includegraphics[width=0.8\linewidth]{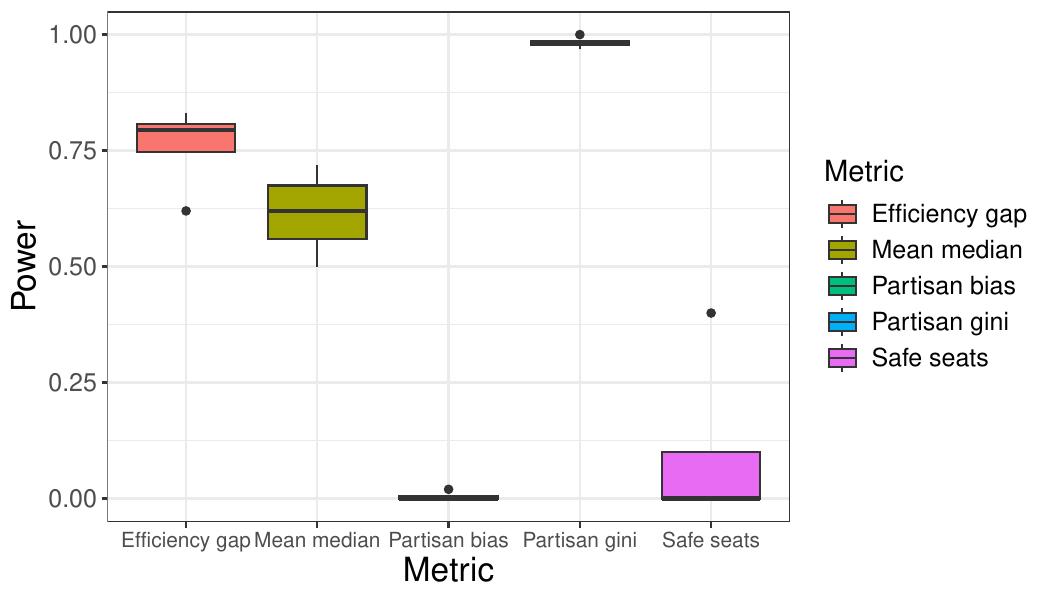}
    \caption{Boxplots of the power for each metric across all $\Sigma^A$ chains biased for that metric. The colors represent the specific metric being tested, for which the chains were also biased.}\label{fig:PowerMetrics}
    \Description[boxplots]{power depends on metrics}
\end{figure}

If an ANOVA is run on all the variables $\omega, P, y$, the only variable which has an effect on power is the different metrics ($\omega$). On average, maps which are biased for partisan gini are the easiest to detect as an $(\alpha, \varepsilon)-$outlier regardless of election year, party, or test parameters. Maps which are biased for efficiency gap and mean median are also easier to detect as $(\alpha, \varepsilon)-$outliers. These results can be seen in Figure~\ref{fig:PowerMetrics} and in the table above.

One important consideration is that the hypothesis test is built for a label $\omega$ which is a real-valued function. Safe seats and partisan bias can only take on a certain number of discrete values making the test harder to use on these metrics and may be the reason why these metrics have lower power on average. 

\subsubsection{Election Year: 2012, 2016}
\begin{figure}[H]
\centering
\includegraphics[width=0.8\linewidth]{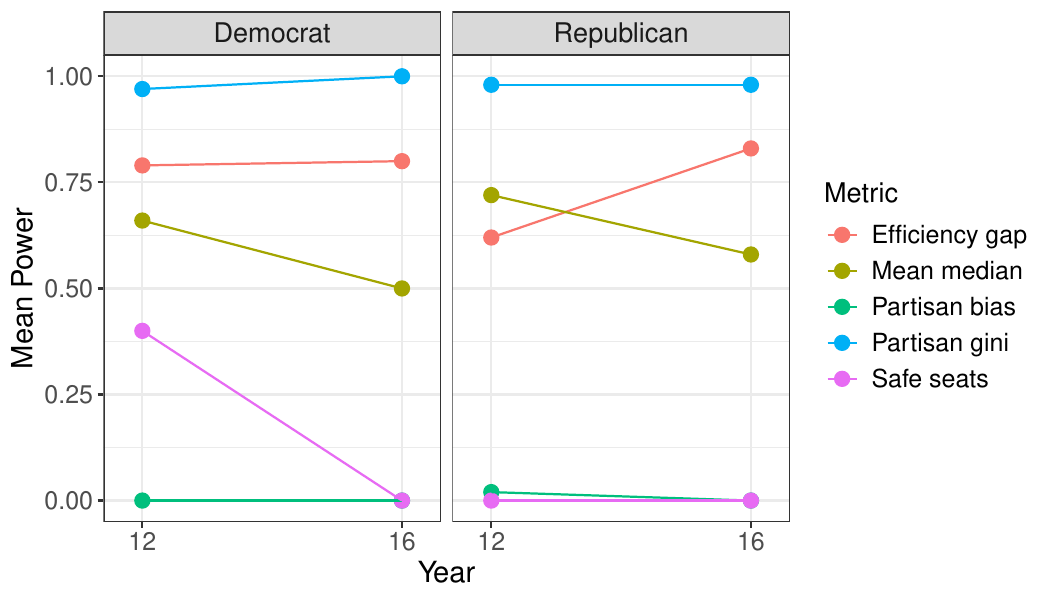}
    \caption{Interaction plots for metric (color), Year (x-axis), and political party with Democrat (left) and Republican (right).}\label{fig:Interactionplots}
    \Description[interaction plots]{no interaction}
\end{figure}

There is no statistically significant difference for election year and political party; there were no interaction between any of the variables and parameters. In general there are small changes the power of certain metrics from the election years: safe seats for democrats was more powerful in 2012 than 2016 as seen in the interaction plots in Figures~\ref{fig:Interactionplots}. This causes the year 2012 to have slightly higher power than for the year 2016. As stated above, safe seats is a harder metric to test since $\omega$ will not be a real-values number. In general, the power did not very that much in each year as seen in the Table below.

\begin{center}
\begin{tabular}{c c} 
\toprule
 \textbf{2012} & \textbf{2016}\\ \midrule
 $0.52 (\pm 0.009)$ & $0.47 (\pm 0.010)$ \\ \bottomrule
\end{tabular}
\end{center}

\subsubsection{Party: Republican and Democrat}

In general, party does not have a huge effect on the power of the test. The republican party has slightly lower power; meaning that a map which is biased for a particular $\omega$ to be gerrymandered to favor Republicans has a slightly lower chance of being detected than the Democrats for these simulations, but this result is not considered to be statistically significant.

\begin{center}
\begin{tabular}{cc} 
 \toprule
 \textbf{Republican} & \textbf{Democrat} \\ \midrule
 $0.49 (\pm 0.04)$ & $0.52 (\pm 0.04)$ \\ \bottomrule
\end{tabular}
\end{center}

In general, metric seems to have the largest effect on power and other variables did not make any noteworthy changes to the power of a given metric.

\section{Discussion}
Though several hypothesis tests have been built for testing a gerrymandering map based on a Markov Chain sampling scheme, relatively little has been done to evaluate their statistical power. Power is crucial for determining the credibility of a test, especially when the procedure is computationally intensive. These tests have been used in several real world court cases, further illustrating the need to validate this method of hypothesis testing. 

The results show that the primary factor which influences the power of these tests is the partisan metric used as the label function $\omega$. When this test is used in the future to access gerrymandering, careful consideration of the power as well as interpretation of the metric used is needed. Other variables which could effect power but were not evaluated in this paper are US state, election type, and different years. 

Another notable finding is that the number of steps in the chains ($k$) required to achieve maximal power was significantly lower than expected. In contrast to typical applications of this test where $k$ can be millions of steps, this analysis suggests that a much smaller $k$ could be used to achieve the same power while significantly reducing the computational cost. A caveat of these results is that the maps used are substantially more biased then real world maps which have been gerrymandered.

\section{Acknowledgements}
This material is based upon work supported by the National Science
Foundation under Grant No. DMS-1916439, while the authors were at AMS Mathematics Research Communities during the summer of 2022 and at Johns Hopkins University during the winter of 2022. It is also based upon work supported by the National Science Foundation under Grant No. DMS-1928930, while the authors were in residence at the Simons Laufer Mathematical Sciences Institute (SLMath) in Berkeley, California, during the summer of 2024. R. A. Clark would also like to acknowledge the support of the NSF MSP Ascending Postdoc Award 2138110.

\bibliographystyle{plain}
\bibliography{biblio}

\end{document}